\documentclass[aps,prl,amsmath,showpacs,preprintnumbers,twocolumn,amssymb,amsmath,superscriptaddress]{revtex4}
\usepackage{graphicx}
\usepackage{mathptmx, textcomp}
\usepackage[latin1]{inputenc}
\usepackage{tabularx}
\usepackage{units}
\usepackage{color}
\usepackage{amsmath}
\usepackage{amssymb}

\begin{document}
\title{Anisotropic relaxation dynamics in a dipolar Fermi gas driven out of equilibrium}

\author{K. Aikawa}
\affiliation{Institut f\"ur Experimentalphysik and Zentrum f\"ur Quantenphysik, Universit\"at Innsbruck, Technikerstra{\ss}e 25, 6020 Innsbruck, Austria}
\author{A. Frisch}
\affiliation{Institut f\"ur Experimentalphysik and Zentrum f\"ur Quantenphysik, Universit\"at Innsbruck, Technikerstra{\ss}e 25, 6020 Innsbruck, Austria}
\author{M. Mark}
\affiliation{Institut f\"ur Experimentalphysik and Zentrum f\"ur Quantenphysik, Universit\"at Innsbruck, Technikerstra{\ss}e 25, 6020 Innsbruck, Austria}
\author{S. Baier}
\affiliation{Institut f\"ur Experimentalphysik and Zentrum f\"ur Quantenphysik, Universit\"at Innsbruck, Technikerstra{\ss}e 25, 6020 Innsbruck, Austria}
\author{R. Grimm}
\affiliation{Institut f\"ur Experimentalphysik and Zentrum f\"ur Quantenphysik, Universit\"at Innsbruck, Technikerstra{\ss}e 25, 6020 Innsbruck, Austria}
\affiliation{Institut f\"ur Quantenoptik und Quanteninformation,
 \"Osterreichische Akademie der Wissenschaften, 6020 Innsbruck, Austria}
\author{J. L. Bohn}
\affiliation{JILA, NIST, and Department of Physics,University of Colorado, Boulder, Colorado 80309, USA}
\author{D. S. Jin}
\affiliation{JILA, NIST, and Department of Physics,University of Colorado, Boulder, Colorado 80309, USA}
\author{G. M. Bruun}
\affiliation{Department of Physics and Astronomy, University of Aarhus, Ny Munkegade, DK-8000 Aarhus C, Denmark}
\author{F. Ferlaino}
\affiliation{Institut f\"ur Experimentalphysik and Zentrum f\"ur Quantenphysik, Universit\"at Innsbruck, Technikerstra{\ss}e 25, 6020 Innsbruck, Austria}
\affiliation{Institut f\"ur Quantenoptik und Quanteninformation,
 \"Osterreichische Akademie der Wissenschaften, 6020 Innsbruck, Austria}

\date{\today}


\begin{abstract}
We report on the observation of a large anisotropy in the rethermalization dynamics of an ultracold dipolar Fermi gas driven out of equilibrium.  Our system consists of an ultracold sample of strongly magnetic $^{167}$Er fermions,  spin-polarized in the lowest Zeeman sublevel. In this system,  elastic collisions arise purely from universal dipolar scattering. Based on cross-dimensional rethermalization experiments, we observe a strong anisotropy of the scattering, which manifests itself in a large  angular dependence of the  thermal relaxation dynamics. Our result is in good agreement with recent theoretical predictions. Furthermore, we measure the rethermalization rate as a function of temperature for different angles and find that the suppression of collisions by Pauli blocking is not influenced by the dipole orientation.

\end{abstract}

\maketitle

The behavior of any many-body system follows from the interactions of its constituent particles. In some cases of physical interest, importantly at ultralow temperature, where the de Broglie wavelength is the dominant length scale, these interactions can be simplified by appealing to the Wigner threshold laws\,\cite{wigner1948behavior,sadeghpour2000collisions}.  These laws, which have been extensively studied for particles interacting via van der Waals forces, both in experiment and theory\,\cite{weiner1999experiments}, identify the interactions via simple isotropic parameters such as a scattering length.  However, for dipolar particles, the fundamental interaction is anisotropic and the system properties can depend on the orientation of the gas with respect to a particular direction in space.\,\cite{lahaye2009thephysics,baranov2012condensed}. 

One of the major strengths of ultracold matter is its susceptibility to being controlled by various means. Striking examples include traversing the BEC-BES crossover\,\cite{varenna2008fermion,giorgini2008theory}, welding atoms together into molecules\,\cite{carr2009cold,jin2012introduction}, and inducing bosons to behave like fermions in one spatial dimension\,\cite{cazalilla2011one,imambekov2012one}.  Most often, such control exploits the quantum mechanical nature of a many-body gas at ultralow temperature, and arises from the manipulation of isotropic scattering between constituent particles.  However, in the case of dipolar particles the scattering is intrinsically anisotropic, affording novel opportunities to control the behavior of the gas.  For example, the anisotropic $d$-wave collapse of a Bose-Einstein condensate of magnetic atoms\,\cite{lahaye2008d-wave,aikawa2012bose-einstein} and the deformation of the Fermi sphere in a dipolar Fermi gas\,\cite{aikawa2014observation} have been observed. These phenomena rely on the collective behavior of all the particles, occurring according to their mean field energy.

Distinct from such many-body effects, dipoles can also influence the properties of the gas via two-body scattering. 
Since scattering of dipoles is highly anisotropic, properties that require the collisional exchange of energy and momentum between the atoms, such as sound propagation, viscosity, and virial coefficients\,\cite{Daily14_PRA}, will be influenced by the presence of dipoles. In particular, differential cross sections of dipolar particles are highly anisotropic, depending on both the initial, as well as scattered, relative directions of the colliding particles, and it has recently been predicted that dipolar anisotropy can exert a profound influence on the non-equilibrium dynamics in such a gas\,\cite{bohn2014differential}.

In this letter, we demonstrate the control of the thermal relaxation dynamics of a dipolar Fermi gas driven out of equilibrium by adding excess momentum along one axis. The control is achieved by changing the orientation of the dipoles relative to this direction, enabling us to substantially vary the rethermaliztion rate. 
%
%
%
As a striking consequence of the interaction anisotropy, we find that the rate of equilibration can vary by as much as a factor of four, depending on the angle between the dipole orientation and the excitation axis. Furthermore, we observe that the rethermalization rate decreases as the temperature is lowered. This effect is due to the lack of available final states into which atoms can scatter and is known as Pauli blocking. Our results provide evidence that the Pauli suppression of collisions does not contribute additional anisotropic effects to the rethermalization.


To realize a dipolar Fermi gas, we use an ultracold spin-polarized sample of strongly magnetic erbium (Er) atoms, which possess a magnetic dipole moment of 7 Bohr magneton. This is an ideal system to study purely dipolar scattering since short-range van der Waals forces give a negligible contribution to the scattering of identical fermions at ultralow temperatures\,\cite{note6}. Our experimental procedure to create a degenerate Fermi gas of  $^{167}$Er atoms follows the one described in Ref.\,\cite{aikawa2014reaching,note7}. In brief,  it  comprises laser cooling in a narrow-line magneto-optical trap (MOT) \cite{frisch2012narrow} and direct evaporative cooling of a spin-polarized sample in an optical dipole trap (ODT)\,\cite{aikawa2014reaching}. 
Evaporative cooling, which was successfully used to reach Bose-Einstein condensation\,\cite{anderson1995observation,davis1995bose}, relies on efficient thermalization. In our case, this  is achieved by elastic dipolar collisions between spin-polarized fermions. 

\begin{figure}[t]
\includegraphics[width=1\columnwidth] {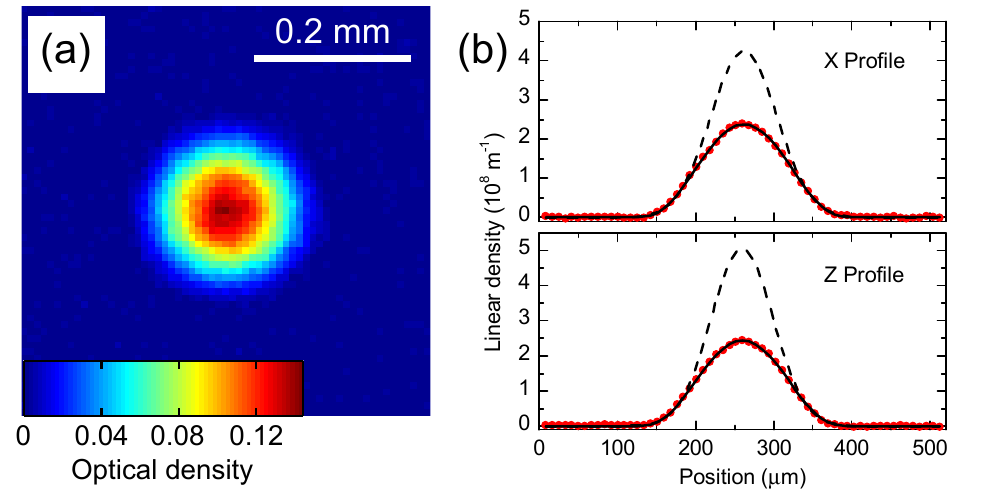}
\caption{(color online) Time-of-flight absorption image of a degenerate Fermi gas of  $3.0 \times 10^4$ Er atoms at $T/T_F = 0.11(1)$ after $\unit[12]{ms}$ of expansion (a) and its density distribution integrated along the $x$ direction (upper panel) and $z$ direction (lower panel) (b). 
The observed profiles (circles) are well described by fitting the Fermi-Dirac distribution to the data (solid lines), while they substantially deviate from a  fit using a Gaussian distribution to the outer wings of the cloud (dashed lines), i.\,e.\,outside the disk with radius $w$, where $w$ is the $1/e$ radius of the Gaussian fit to the the entire cloud. The absorption image is averaged over 20 individual measurements.}
\label{fig:fig_image}
\end{figure}

We optimize the evaporative cooling sequence and produce a degenerate Fermi gas of about $3.0 \times 10^4$ atoms at a temperature as low as $T/T_F = 0.11(1)$. This gives a peak density of about $\unit[3\times10^{14}]{cm^{-3}}$; see Fig.\,\ref{fig:fig_image}. Here, $T_F = \unit[1.06(5)]{\mu K}$. To achieve such a deeply degenerate regime, we confine the atoms more tightly than in our previous work\,\cite{aikawa2014reaching} by decreasing the beam waist of the vertical \unit[1570]{nm} beam from $33$ to $\unit[21]{\mu m}$. The trap frequencies in this configuration are $(\nu_x, \nu_y, \nu_z) = \unit[(509, 447, 262)]{Hz}$. Our minimum temperature is comparable to the lowest ones achieved with sympathetic cooling schemes based on $s$-wave scattering\,\cite{Inguscio2006ufg,taie2010realization,stellmer2013production}.

\begin{figure}[t]
\includegraphics[width=1\columnwidth] {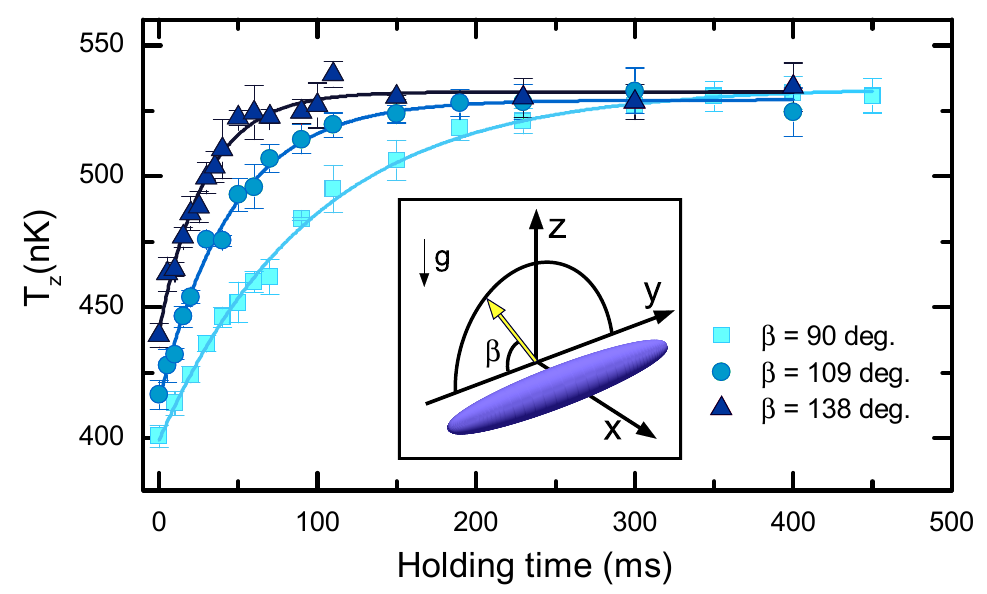}
\caption{ (color online) Typical cross-dimensional thermalization measurements for three dipole orientations: $\beta= \unit[90]{^\circ}$ (squares), $\beta= \unit[109]{^\circ}$ (circles), and $\beta= \unit[138]{^\circ}$ (triangles). The time evolution of the temperature in the $z$ direction, $T_z$, is plotted as a function of holding time after the cloud is excited in the $y$ direction. The geometry of the cigar-shaped trap and the coordinates are indicated in the inset. 
The yellow arrow represents the dipole orientation.
After the equilibration, the Fermi gas is at $T/T_F \simeq 0.75$ with $T_F \simeq \unit[710]{nK}$.}
\label{fig:fig_therm}
\end{figure}

While the crucial role of the long-range character of the dipole-dipole interaction (DDI) clearly emerges in the evaporation of spin-polarized fermions, the role of the anisotropy of the interaction is  more subtle. We investigate this aspect by studying for various dipole orientations how the system rethermalizes when it is excited out of its equilibrium state. The dipole orientation is  controlled by changing the direction of the polarizing magnetic field and is represented by the angle $\beta$ between the magnetic field and the weak axis of the ODT; see inset Fig.\,\ref{fig:fig_therm}. 

The cross-dimensional rethermalization experiments proceed as follows. We first prepare a nearly degenerate Fermi gas of about $8\times 10^4$ atoms at  $T/T_F \simeq 0.6$ with $T_F \simeq \unit[600]{nK}$ in a cigar-shaped ODT with frequencies of $\unit[(393, 23, 418)]{Hz}$. We then change the dipole orientation from $\beta=90^{\circ}$, which is used for evaporative cooling, to the desired value and excite the system by increasing the power of the vertical beam by about a factor of 2.8 within $\unit[14]{ms}$. After the excitation, the trap frequencies are $\unit[(393, 38, 418)]{Hz}$. The excitation brings the system out of equilibrium by transferring energy mainly in the direction of the weak axis ($y$), which we refer to as  the {\em excitation axis}. This process is nearly adiabatic with respect to the trap period for the radial motion and introduces an initial temperature imbalance of about $T_y/T_z=2$. We follow the re-equilibration dynamics by recording  the time evolution of the temperature along the $z$ direction, $T_z$, and we extract the rethermalization rate for the given dipole orientation.

Figure\,\ref{fig:fig_therm} shows typical temperature evolutions, measured for three different values of $\beta$. We observe that the re-equilibration dynamics strongly depends on $\beta$ with $T_z$  approaching the new equilibrium value in a near-exponential way. From an exponential fit to the data, we extract  the  relaxation time constant $\tau$. The slower rethermalization  (i.\,e.\,largest $\tau$) is found to occur  for dipoles oriented perpendicular to the excitation axis, whereas by changing $\beta$ by about $\unit[45]{^\circ}$ we observe that the system re-equilibrates about four times faster. This strong angle dependence clearly shows the anisotropic nature of the relaxation dynamics.

Since rethermalization relies on elastic collisions, its rate $1/\tau$ should be directly proportional to the total elastic cross section $\sigma_{\rm el}$\,\cite{bohn2014differential}.  
The standard way to evaluate the latter is  by integrating the differential cross sections $d \sigma_{\rm el} /d \Omega ({\bf k}^{\prime},{\bf k})$ over all final directions of the atoms' relative momentum ${\bf k}^{\prime}$, and averaging over incident directions ${\bf k}$. For  dipolar fermions, the integration yields an energy-independent $\sigma_{\rm el}$, which is universally related to the fermions' dipole moment $d$ by $\sigma_{\rm el}= (32 \pi /15)D^2$, where $D=2 \pi^2 d^2 m / h^2$, with $m$ the mass and $h$ the Planck constant, is the characteristic length scale of the dipolar interaction \cite{bohn2009quasi-universal}.
A collision rate can then be defined as ${\bar n}  { \sigma}_{el} v$, where ${\bar n}$ is the mean number density and $v=\sqrt{16 k_B T'/(\pi m)}$ is the mean relative velocity with $T'$ the effective temperature including a momentum spread by the Fermi energy\,\cite{note4}.

The actual re-equilibration occurs at a rate different from this collision rate since rethermalization emphasizes those collisions that significantly change the relative direction of the atoms' momenta. The characteristic time for the relaxation is inversely proportional to the collision rate
\begin{eqnarray}
\tau = \frac{ \alpha }{ {\bar n} \eta {\sigma}_{\rm el} v }.
 \label{eq1}
\end{eqnarray}
The dimensionless proportionality constant $\alpha$ is commonly referred to as the number of collisions for thermalization and can be computed from the known differential cross sections.
For short-range interactions, Monte-Carlo calculations yield $\alpha=2.7$  and $\alpha=4.1$ for $s$- and $p$-wave collisions, respectively~\cite{monroe1993measurement,demarco1999measurement}.
As predicted in Ref.\,\cite{bohn2014differential}, in the case of the DDI, which is  long-range and  anisotropic, $\alpha$ is a function of the angle $\beta$. In Eq.(\ref{eq1}), we also include a Pauli suppression factor, $\eta$, which accounts for the  reduction of the rethermalization rate in a degenerate Fermi gas caused by Pauli blocking  where $\eta=1$ for nondegenerate gases; see later discussion.

\begin{figure}[t]
\includegraphics[width=1\columnwidth] {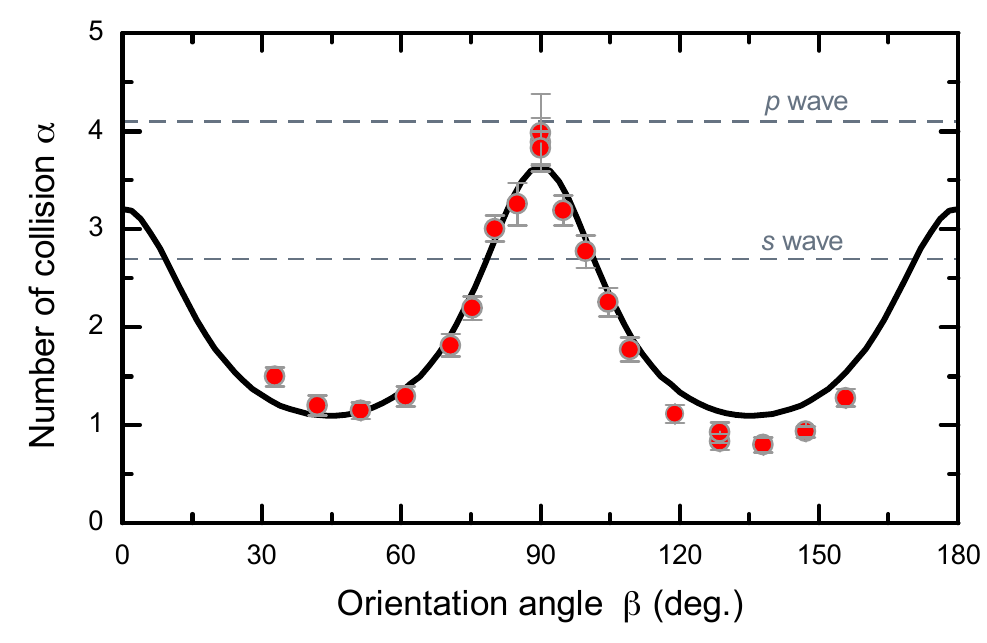}
\caption{(color online) Angle dependence of the number of collisions, $\alpha$, required to rethermalize ultracold dipolar fermions.  The experimental data (circles) are compared with the parameter-free theoretical prediction calculated under our trapping and excitation conditions (solid line).
}
\label{fig:fig_angledep}
\end{figure}

To explore the angle dependence of $\alpha$, we perform cross-dimensional rethermalization experiments under the same conditions as in Fig.\,\ref{fig:fig_therm} for various values of $\beta$ in a range between $30^\circ$ and $160^\circ$. We extract $\alpha$ by using Eq.\,(\ref{eq1}),  the experimentally measured $\tau$, and the elastic dipolar cross-section $\sigma_{\rm el}= 1.8 \times 10^{-12}$ cm$^2$ calculated for Er\,\cite{bohn2009quasi-universal}. At $T/T_F\simeq 0.75$, which is our experimental conditions after excitation,  $\eta=0.93$; see later discussion.  
Figure\,\ref{fig:fig_angledep} shows our experimental result together with the theoretical value of $\alpha$, which we calculate using Enskog's equation similarly to Ref.\,\cite{bohn2014differential}.  We observe a remarkably good  quantitative agreement between the experiment and the theory, which does not have any free parameter. The theoretical curve is calculated assuming a
 velocity distribution that is Maxwell-Boltzmann in form that allows the width of the distribution to be characterized by different ``temperatures'' $T_x$, $T_y$, and $T_z$ in the three directions of the trap. We then use the collision integral in Enskog's formulation to calculate the relaxation rate for $T_z$ for the following initial conditions: $T_z = T_x$, and $T_y = 2T_z$. From this we determine the theoretical value of $\alpha$ along the $z$ direction; see Fig.\,\ref{fig:fig_angledep}.  Owing to the anisotropy of the dipolar collision cross section, we find that $\alpha$ is, in general, different in the three directions $i=x,y,z$. 

Both the experimental and theoretical results show a strong angular dependence of $\alpha$, which largely varies in a sine-like manner from about 1 to 4. This behavior is unique to dipolar scattering  and  occurs because dipolar particles have a spatial orientation dictated by the magnetic field and the collision processes depend on this orientation \cite{bohn2014differential}. Given this angular dependence, dipoles can re-equilibrate on a timescale that can be even faster than the one achievable with short-range $s$-wave collisions ($\alpha=2.7$).  In particular,  the rethermalization is the most efficient (smallest $\alpha$) when $\beta = \unit[45]{^\circ}$.  This can be seen qualitatively from the form of the cross sections in Ref.\,\cite{bohn2014differential}.  Consider a simplified scattering event, where the magnetic field ${\bf B}$, the relative incident  wave vector ${\bf k}$, and the relative scattered wave vector ${\bf k}^{\prime}$ all lie in the same plane.  Moreover, suppose that ${\bf k}$ makes an angle $\theta$ with respect to ${\bf B}$, while ${\bf k}^{\prime}$ makes an angle $\theta^{\prime}$, defining a scattering angle $\theta_s = \theta^{\prime} - \theta$.  In this reduced case, the differential cross section for scattering of dipolar fermions has the simple angular dependence $d \sigma_{\rm el} / d \Omega \propto \cos^2(2 \theta -  \theta_s)$, thus the most likely scattering occurs when $\theta_s = 2\theta$.  For the most efficient rethermalization one requires scattering at right angles, $\theta_s = \unit[90]{^\circ}$, and therefore collisions in which the atoms approach one another at $\theta = \unit[45]{^\circ}$.  At the same time, from the experimental geometry one requires a high collision rate between the radial and axial trap directions.  From these considerations, the most efficient rethermalization should occur for a field tilted at an angle $\beta = \unit[45]{^\circ}$ from the trap's axis.  

\begin{figure}[t]
\includegraphics[width=1\columnwidth] {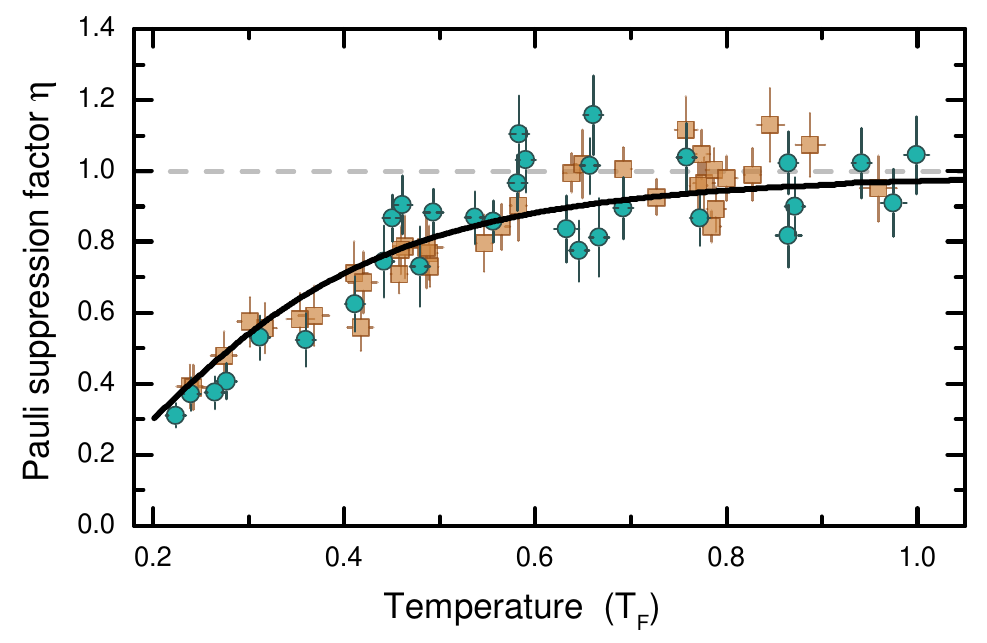}
\caption{(color online) The Pauli suppression factor $\eta$ as a function of the temperature. 
The experimental data are taken for two different orientations of the dipoles, i.\,e.\,for $\beta = \unit[90]{^\circ}$ (squares) and for $\beta=\unit[110]{^\circ}$ (circles). The experimental values are compared with the theoretical predictions on Pauli blocking (solid line); see text.}
\label{fig:fig_tempdep}
\end{figure}

Another central aspect in the scattering of ultracold fermions in the degenerate regime is related to Pauli blocking of collisions. This effect is caused by the lack of unoccupied final states in a scattering event and leads to a reduction of the elastic collision rate, which has been observed in fermionic systems with short-range interaction\,\cite{demarco1999measurement,Riedl2008coo}. In the case of dipolar scattering, it is an interesting question whether or not the Pauli blocking effect  exhibits anisotropy. 

To address this question experimentally,  we explore the temperature dependences of $\tau$ for two different angles, $\beta = \unit[90]{^\circ}$ and $\unit[110]{^\circ}$, for which $\alpha$ differs by a factor of two; see Fig.\,\ref{fig:fig_angledep}.  For each temperature, we measure $\tau$, $n$, $v$ and we derive $\eta/\alpha$ using Eq.\,(\ref{eq1}). To extract the Pauli suppression factor $\eta$, we normalize $\eta/\alpha$ to  its value in the non-degenerate regime where $\eta=1$.  As shown in Fig.\,\ref{fig:fig_tempdep}, for both angles we observe the same pronounced reduction of $\eta$ with decreasing  temperature. Specifically, $\eta$ decreases by about $70\%$  from $1$ to $0.2 \,T/T_F$. Our results  indicate that the reduction of the rethermalization rate is independent from the dipole orientation.

In order to quantitatively confirm that the observed reduction of $\eta$ is caused by Pauli blocking of collisions, we use a theoretical model following the one developed for non-dipolar fermions in Ref.\,\cite{Massignan2005vra,Bruun2005vat}.  Our model is based on a variational approach for solving the  Boltzmann equation to calculate the thermal relaxation rate including the Pauli blocking effect. 
This gives~\cite{Pethick2002book}
 \begin{equation}
\frac 1 {\tau}=\frac{\langle\Gamma[\Phi_T]\Phi_T\rangle}{2\langle\Phi_T^2\rangle}
\label{Rate}
\end{equation}
where the trap and momentum average is defined as $\langle\ldots\rangle=\int d^3r\int d^3\check k f(1-f)\ldots$ with $d^3\check k =d^3k/(2\pi)^3 $, and 
$f(k,{\mathbf r})=\{\exp[\beta(k^2/2m+V({\mathbf r})-\mu)]+1\}^{-1}$
is the equilibrium Fermi function with $V({\mathbf r})$ the trapping potential.  We have defined $\Gamma[\Phi_T]=I[\Phi_T]/f(1-f)$ with 
\begin{equation}
 I[\Phi]=\int d^3\check k_2 d\Omega\frac{d\sigma_{\rm el}}{d\Omega}\frac{|{\mathbf k}-{\mathbf k}_2|}{m}\Delta\Phi
ff_2(1-f_3)(1-f_4)
\label{CollisionOperator}
 \end{equation}
the collision integral describing the rethermalization due to the collision of two  particles with 
incoming momenta ${\mathbf k}$ and ${\mathbf k}_2$ and outgoing  momenta ${\mathbf k}_3$ and ${\mathbf k}_4$.  Here, 
 $\Delta\Phi=\Phi_1+\Phi_2-\Phi_3-\Phi_4$, and $\Omega$ is the solid angle of the outgoing 
relative momentum $({\mathbf k}_4-{\mathbf k}_3)/2$. 
We use $\Phi_T=k_y^2-k^2/3$ as a variational expression for the  deviation function corresponding to a thermal  anisotropy 
in the $y$-direction. This is also the approach used to calculate $\alpha$ as a function of $\beta$ for a Maxwell-Boltzmann distribution in Ref.\,\cite{bohn2014differential} and used to calculated the shear viscosity in a gas interacting with $s$-wave interactions~\cite{Massignan2005vra,Bruun2005vat}.

By calculating  the thermal relaxation rate using Eq.\,(\ref{Rate}) with either a pure $s$-wave or a pure $p$-wave interaction, we find that the relative suppression of the rate compared to the classical value  is essentially the same in the two partial wave channels for a given temperature.
This indicates that the Pauli blocking effect is largely insensitive to the angular dependence of the cross section of the atoms. Hence, we can safely use the $p$-wave cross section as a stand-in for the true dipolar scattering in this calculation. In the high-temperature limit  $T\gg T_F$, the integrals in Eqs.\ (\ref{Rate})-(\ref{CollisionOperator}) can be solved analytically and we get  Eq.\ (\ref{eq1}) with $\alpha=25/6$ for $p$-wave scattering and $\alpha= 5/2$ for $s$-wave scattering. These variational values agree well with the Monte-Carlo results stated above.

 In Fig.\,\ref{fig:fig_tempdep}, we plot the theoretically predicted temperature dependence of $\eta$ using Eqs.\ (\ref{Rate}) and (\ref{CollisionOperator}) and assuming a $p$-wave cross section, together with the experimental values. As explained above, the calculation yields essentially the same result when we assume  $s$-wave scattering.
 By fitting the theoretical curve to the observed rethermalization rates with a single free scaling parameter for each angle, we determine the values of $\sigma_{\rm el}/\alpha$ to be $\unit[0.51(1)(12)\times 10^{-12}]{cm^2}$ at $\beta = \unit[90]{^\circ}$ and $\unit[1.00(2)(24)\times 10^{-12}]{cm^2}$ at $\beta = \unit[110]{^\circ}$,  where the errors are statistical errors of fitting and systematic uncertainties on temperature, trap frequencies, and atom number, respectively. 
The theory and experiment results are  in good agreement. Our findings show that the reduction of the rethermalization rate is indeed due to Pauli blocking and that it is not significantly influenced by the polarization angle.

In conclusion, we have studied the rethermalization dynamics of indistinguishable dipolar fermions after the system is excited out of thermal equilibrium.  We have observed that the rate of equilibration can strongly vary  depending on the polarization direction of the atoms' magnetic dipole moments, which demonstrates a remarkable influence of anisotropic scattering.  Further, we note that the anisotropy has no significant influence on the Pauli blocking effect.  Our results are fundamentally important for understanding the collisional behavior of dipolar particles, such as strongly magnetic atoms and polar molecules.

\begin{acknowledgments}
We are grateful to M.\,Baranov for fruitful discussions. This work is supported by the Austrian Ministry of Science and Research (BMWF) and the Austrian Science Fund (FWF) through a START grant under Project No.\,Y479-N20 and by the European Research Council under Project No.\,259435. K.\,A.\,is supported within the Lise-Meitner program of the FWF. J.\,L.\,B.\,and D.\,S.\,J.\,acknowledge support from U.S. National Science Foundation grant number 1125844.
\end{acknowledgments}

\appendix

\section{Supplementary Material}

\subsection{Preparing spin-polarized samples of $^{167}$Er}
After the MOT stage, in which the sample is simultaneously cooled and spin-polarized into the lowest hyperfine sublevel,  we typically transfer $1.5 \times 10^6$ atoms  at a temperature of $T = \unit[28]{\mu K} $ into a crossed ODT. The latter consists of a horizontal beam, which propagates along the $y$ direction, and a vertical beam, propagating along the $z$ direction (direction of gravity). The horizontal beam has a beam waist of $\unit[15]{\mu m}$ and a wavelength of $\unit[1570]{nm}$. For the vertical beam, we use light that operates either at  $\unit[1570]{nm}$ or at $\unit[1064]{nm}$ and various values of the beam waist, depending on the final temperatures and densities required for the specific experiments. 

\subsection{Magnetic field control}
 We apply a magnetic bias field of $\unit[0.58]{G}$, oriented along the $z$ direction. We select this value of the magnetic field to avoid loss features associated to the recently observed Fano-Feshbach resonances\,\cite{frisch2014quantum}. At $\unit[0.58]{G}$, the Zeeman energy splitting between magnetic sublevels is $\unit[0.6]{MHz}\times h$. This value is much larger than the typical Fermi energy (about $\unit[20]{kHz}\times h$) and the typical energy of the DDI (about $\unit[100]{Hz}\times h$).  

The angle $\beta$ and the amplitude of the magnetic field are controlled by three independent sets of coils along the  $x$,  $y$, and $z$ directions. Each coil set is independently calibrated by using radio-frequency spectroscopy to within $\unit[5]{mG}$, from which we estimate the error on angle to be within 1 degree.

\bibliographystyle{apsrev}


\begin{thebibliography}{34}
\expandafter\ifx\csname natexlab\endcsname\relax\def\natexlab#1{#1}\fi
\expandafter\ifx\csname bibnamefont\endcsname\relax
  \def\bibnamefont#1{#1}\fi
\expandafter\ifx\csname bibfnamefont\endcsname\relax
  \def\bibfnamefont#1{#1}\fi
\expandafter\ifx\csname citenamefont\endcsname\relax
  \def\citenamefont#1{#1}\fi
\expandafter\ifx\csname url\endcsname\relax
  \def\url#1{\texttt{#1}}\fi
\expandafter\ifx\csname urlprefix\endcsname\relax\def\urlprefix{URL }\fi
\providecommand{\bibinfo}[2]{#2}
\providecommand{\eprint}[2][]{\url{#2}}

\bibitem[{\citenamefont{Wigner}(1948)}]{wigner1948behavior}
\bibinfo{author}{\bibfnamefont{E.~P.} \bibnamefont{Wigner}},
  \bibinfo{journal}{Phys. Rev.} \textbf{\bibinfo{volume}{73}},
  \bibinfo{pages}{1002} (\bibinfo{year}{1948}).

\bibitem[{\citenamefont{Sadeghpour et~al.}(2000)\citenamefont{Sadeghpour, Bohn,
  Cavagnero, Esry, Fabrikant, Macek, and Rau}}]{sadeghpour2000collisions}
\bibinfo{author}{\bibfnamefont{H.}~\bibnamefont{Sadeghpour}},
  \bibinfo{author}{\bibfnamefont{J.}~\bibnamefont{Bohn}},
  \bibinfo{author}{\bibfnamefont{M.}~\bibnamefont{Cavagnero}},
  \bibinfo{author}{\bibfnamefont{B.}~\bibnamefont{Esry}},
  \bibinfo{author}{\bibfnamefont{I.}~\bibnamefont{Fabrikant}},
  \bibinfo{author}{\bibfnamefont{J.}~\bibnamefont{Macek}}, \bibnamefont{and}
  \bibinfo{author}{\bibfnamefont{A.}~\bibnamefont{Rau}}, \bibinfo{journal}{J.
  Phys. B} \textbf{\bibinfo{volume}{33}}, \bibinfo{pages}{R93}
  (\bibinfo{year}{2000}).

\bibitem[{\citenamefont{Weiner et~al.}(1999)\citenamefont{Weiner, Bagnato,
  Zilio, and Julienne}}]{weiner1999experiments}
\bibinfo{author}{\bibfnamefont{J.}~\bibnamefont{Weiner}},
  \bibinfo{author}{\bibfnamefont{V.~S.} \bibnamefont{Bagnato}},
  \bibinfo{author}{\bibfnamefont{S.}~\bibnamefont{Zilio}}, \bibnamefont{and}
  \bibinfo{author}{\bibfnamefont{P.~S.} \bibnamefont{Julienne}},
  \bibinfo{journal}{Rev. Mod. Phys.} \textbf{\bibinfo{volume}{71}},
  \bibinfo{pages}{1} (\bibinfo{year}{1999}).

\bibitem[{\citenamefont{Lahaye et~al.}(2009)\citenamefont{Lahaye, Menotti,
  Santos, Lewenstein, and Pfau}}]{lahaye2009thephysics}
\bibinfo{author}{\bibfnamefont{T.}~\bibnamefont{Lahaye}},
  \bibinfo{author}{\bibfnamefont{C.}~\bibnamefont{Menotti}},
  \bibinfo{author}{\bibfnamefont{L.}~\bibnamefont{Santos}},
  \bibinfo{author}{\bibfnamefont{M.}~\bibnamefont{Lewenstein}},
  \bibnamefont{and} \bibinfo{author}{\bibfnamefont{T.}~\bibnamefont{Pfau}},
  \bibinfo{journal}{Rep. Prog. Phys.} \textbf{\bibinfo{volume}{72}},
  \bibinfo{pages}{126401} (\bibinfo{year}{2009}).

\bibitem[{\citenamefont{Baranov et~al.}(2012)\citenamefont{Baranov, Dalmonte,
  Pupillo, and Zoller}}]{baranov2012condensed}
\bibinfo{author}{\bibfnamefont{M.~A.} \bibnamefont{Baranov}},
  \bibinfo{author}{\bibfnamefont{M.}~\bibnamefont{Dalmonte}},
  \bibinfo{author}{\bibfnamefont{G.}~\bibnamefont{Pupillo}}, \bibnamefont{and}
  \bibinfo{author}{\bibfnamefont{P.}~\bibnamefont{Zoller}},
  \bibinfo{journal}{Chem. Rev.} \textbf{\bibinfo{volume}{112}},
  \bibinfo{pages}{5012} (\bibinfo{year}{2012}).

\bibitem[{var(2008)}]{varenna2008fermion}
\emph{\bibinfo{title}{Ultra-cold Fermi Gases: Proceedings of the International
  School of Physics "Enrico Fermi", Course CLXIV}}
  (\bibinfo{address}{Amsterdam}, \bibinfo{year}{2008}), \bibinfo{note}{varenna,
  20 to 30 June 2006}.

\bibitem[{\citenamefont{Giorgini et~al.}(2008)\citenamefont{Giorgini,
  Pitaevskii, and Stringari}}]{giorgini2008theory}
\bibinfo{author}{\bibfnamefont{S.}~\bibnamefont{Giorgini}},
  \bibinfo{author}{\bibfnamefont{L.~P.} \bibnamefont{Pitaevskii}},
  \bibnamefont{and}
  \bibinfo{author}{\bibfnamefont{S.}~\bibnamefont{Stringari}},
  \bibinfo{journal}{Rev. Mod. Phys.} \textbf{\bibinfo{volume}{80}},
  \bibinfo{pages}{1215} (\bibinfo{year}{2008}).

\bibitem[{\citenamefont{Carr et~al.}(2009)\citenamefont{Carr, DeMille, Krems,
  and Ye}}]{carr2009cold}
\bibinfo{author}{\bibfnamefont{L.~D.} \bibnamefont{Carr}},
  \bibinfo{author}{\bibfnamefont{D.}~\bibnamefont{DeMille}},
  \bibinfo{author}{\bibfnamefont{R.~V.} \bibnamefont{Krems}}, \bibnamefont{and}
  \bibinfo{author}{\bibfnamefont{J.}~\bibnamefont{Ye}}, \bibinfo{journal}{New
  J. Phys.} \textbf{\bibinfo{volume}{11}}, \bibinfo{pages}{055049}
  (\bibinfo{year}{2009}).

\bibitem[{\citenamefont{Jin and Ye}(2012)}]{jin2012introduction}
\bibinfo{author}{\bibfnamefont{D.~S.} \bibnamefont{Jin}} \bibnamefont{and}
  \bibinfo{author}{\bibfnamefont{J.}~\bibnamefont{Ye}}, \bibinfo{journal}{Chem.
  Rev.} \textbf{\bibinfo{volume}{112}}, \bibinfo{pages}{4801}
  (\bibinfo{year}{2012}).

\bibitem[{\citenamefont{Cazalilla et~al.}(2011)\citenamefont{Cazalilla, Citro,
  Giamarchi, Orignac, and Rigol}}]{cazalilla2011one}
\bibinfo{author}{\bibfnamefont{M.~A.} \bibnamefont{Cazalilla}},
  \bibinfo{author}{\bibfnamefont{R.}~\bibnamefont{Citro}},
  \bibinfo{author}{\bibfnamefont{T.}~\bibnamefont{Giamarchi}},
  \bibinfo{author}{\bibfnamefont{E.}~\bibnamefont{Orignac}}, \bibnamefont{and}
  \bibinfo{author}{\bibfnamefont{M.}~\bibnamefont{Rigol}},
  \bibinfo{journal}{Rev. Mod. Phys.} \textbf{\bibinfo{volume}{83}},
  \bibinfo{pages}{1405} (\bibinfo{year}{2011}).

\bibitem[{\citenamefont{Imambekov et~al.}(2012)\citenamefont{Imambekov,
  Schmidt, and Glazman}}]{imambekov2012one}
\bibinfo{author}{\bibfnamefont{A.}~\bibnamefont{Imambekov}},
  \bibinfo{author}{\bibfnamefont{T.~L.} \bibnamefont{Schmidt}},
  \bibnamefont{and} \bibinfo{author}{\bibfnamefont{L.~I.}
  \bibnamefont{Glazman}}, \bibinfo{journal}{Rev. Mod. Phys.}
  \textbf{\bibinfo{volume}{84}}, \bibinfo{pages}{1253} (\bibinfo{year}{2012}).

\bibitem[{\citenamefont{Lahaye et~al.}(2008)\citenamefont{Lahaye, Metz,
  Froehlich, Koch, Meister, Griesmaier, Pfau, Saito, Kawaguchi, and
  Ueda}}]{lahaye2008d-wave}
\bibinfo{author}{\bibfnamefont{T.}~\bibnamefont{Lahaye}},
  \bibinfo{author}{\bibfnamefont{J.}~\bibnamefont{Metz}},
  \bibinfo{author}{\bibfnamefont{B.}~\bibnamefont{Froehlich}},
  \bibinfo{author}{\bibfnamefont{T.}~\bibnamefont{Koch}},
  \bibinfo{author}{\bibfnamefont{M.}~\bibnamefont{Meister}},
  \bibinfo{author}{\bibfnamefont{A.}~\bibnamefont{Griesmaier}},
  \bibinfo{author}{\bibfnamefont{T.}~\bibnamefont{Pfau}},
  \bibinfo{author}{\bibfnamefont{H.}~\bibnamefont{Saito}},
  \bibinfo{author}{\bibfnamefont{Y.}~\bibnamefont{Kawaguchi}},
  \bibnamefont{and} \bibinfo{author}{\bibfnamefont{M.}~\bibnamefont{Ueda}},
  \bibinfo{journal}{Phys. Rev. Lett.} \textbf{\bibinfo{volume}{101}},
  \bibinfo{pages}{080401} (\bibinfo{year}{2008}).

\bibitem[{\citenamefont{Aikawa et~al.}(2012)\citenamefont{Aikawa, Frisch, Mark,
  Baier, Rietzler, Grimm, and Ferlaino}}]{aikawa2012bose-einstein}
\bibinfo{author}{\bibfnamefont{K.}~\bibnamefont{Aikawa}},
  \bibinfo{author}{\bibfnamefont{A.}~\bibnamefont{Frisch}},
  \bibinfo{author}{\bibfnamefont{M.}~\bibnamefont{Mark}},
  \bibinfo{author}{\bibfnamefont{S.}~\bibnamefont{Baier}},
  \bibinfo{author}{\bibfnamefont{A.}~\bibnamefont{Rietzler}},
  \bibinfo{author}{\bibfnamefont{R.}~\bibnamefont{Grimm}}, \bibnamefont{and}
  \bibinfo{author}{\bibfnamefont{F.}~\bibnamefont{Ferlaino}},
  \bibinfo{journal}{Phys. Rev. Lett.} \textbf{\bibinfo{volume}{108}},
  \bibinfo{pages}{210401} (\bibinfo{year}{2012}).

\bibitem[{\citenamefont{Aikawa et~al.}(2014{\natexlab{a}})\citenamefont{Aikawa,
  Baier, Frisch, Mark, Ravensbergen, and Ferlaino}}]{aikawa2014observation}
\bibinfo{author}{\bibfnamefont{K.}~\bibnamefont{Aikawa}},
  \bibinfo{author}{\bibfnamefont{S.}~\bibnamefont{Baier}},
  \bibinfo{author}{\bibfnamefont{A.}~\bibnamefont{Frisch}},
  \bibinfo{author}{\bibfnamefont{M.}~\bibnamefont{Mark}},
  \bibinfo{author}{\bibfnamefont{C.}~\bibnamefont{Ravensbergen}},
  \bibnamefont{and} \bibinfo{author}{\bibfnamefont{F.}~\bibnamefont{Ferlaino}},
  \bibinfo{journal}{Science} \textbf{\bibinfo{volume}{345}},
  \bibinfo{pages}{1484} (\bibinfo{year}{2014}{\natexlab{a}}).

\bibitem[{\citenamefont{Daily and Blume}(2014)}]{Daily14_PRA}
\bibinfo{author}{\bibfnamefont{K.~M.} \bibnamefont{Daily}} \bibnamefont{and}
  \bibinfo{author}{\bibfnamefont{D.}~\bibnamefont{Blume}},
  \bibinfo{journal}{Phys. Rev. A} \textbf{\bibinfo{volume}{89}},
  \bibinfo{pages}{013606} (\bibinfo{year}{2014}).

\bibitem[{\citenamefont{Bohn and Jin}(2014)}]{bohn2014differential}
\bibinfo{author}{\bibfnamefont{J.~L.} \bibnamefont{Bohn}} \bibnamefont{and}
  \bibinfo{author}{\bibfnamefont{D.~S.} \bibnamefont{Jin}},
  \bibinfo{journal}{Phys. Rev. A} \textbf{\bibinfo{volume}{89}},
  \bibinfo{pages}{022702} (\bibinfo{year}{2014}).

\bibitem[{not({\natexlab{a}})}]{note6}
\bibinfo{note}{According to the Wigner threshold law for the short-range van
  der Waals interaction, the cross section for identical fermions rapidly
  vanishes with temperature. As a consequence, both the isotropic and
  anisotropic part of the van der Waals interaction plays a negligible role in
  the scattering.}

\bibitem[{\citenamefont{Aikawa et~al.}(2014{\natexlab{b}})\citenamefont{Aikawa,
  Frisch, Mark, Baier, Grimm, and Ferlaino}}]{aikawa2014reaching}
\bibinfo{author}{\bibfnamefont{K.}~\bibnamefont{Aikawa}},
  \bibinfo{author}{\bibfnamefont{A.}~\bibnamefont{Frisch}},
  \bibinfo{author}{\bibfnamefont{M.}~\bibnamefont{Mark}},
  \bibinfo{author}{\bibfnamefont{S.}~\bibnamefont{Baier}},
  \bibinfo{author}{\bibfnamefont{R.}~\bibnamefont{Grimm}}, \bibnamefont{and}
  \bibinfo{author}{\bibfnamefont{F.}~\bibnamefont{Ferlaino}},
  \bibinfo{journal}{Phys. Rev. Lett.} \textbf{\bibinfo{volume}{112}},
  \bibinfo{pages}{010404} (\bibinfo{year}{2014}{\natexlab{b}}).

\bibitem[{not({\natexlab{b}})}]{note7}
\bibinfo{note}{See Supplemental Material for experimental details, which
  includes Ref.\,\cite{frisch2014quantum}.}

\bibitem[{\citenamefont{Frisch et~al.}(2014)\citenamefont{Frisch, Mark, Aikawa,
  Ferlaino, Bohn, Makrides, Petrov, and Kotochigova}}]{frisch2014quantum}
\bibinfo{author}{\bibfnamefont{A.}~\bibnamefont{Frisch}},
  \bibinfo{author}{\bibfnamefont{M.}~\bibnamefont{Mark}},
  \bibinfo{author}{\bibfnamefont{K.}~\bibnamefont{Aikawa}},
  \bibinfo{author}{\bibfnamefont{F.}~\bibnamefont{Ferlaino}},
  \bibinfo{author}{\bibfnamefont{J.~L.} \bibnamefont{Bohn}},
  \bibinfo{author}{\bibfnamefont{C.}~\bibnamefont{Makrides}},
  \bibinfo{author}{\bibfnamefont{A.}~\bibnamefont{Petrov}}, \bibnamefont{and}
  \bibinfo{author}{\bibfnamefont{S.}~\bibnamefont{Kotochigova}},
  \bibinfo{journal}{Nature} \textbf{\bibinfo{volume}{507}},
  \bibinfo{pages}{475} (\bibinfo{year}{2014}).

\bibitem[{\citenamefont{Frisch et~al.}(2012)\citenamefont{Frisch, Aikawa, Mark,
  Rietzler, Schindler, Zupanic, Grimm, and Ferlaino}}]{frisch2012narrow}
\bibinfo{author}{\bibfnamefont{A.}~\bibnamefont{Frisch}},
  \bibinfo{author}{\bibfnamefont{K.}~\bibnamefont{Aikawa}},
  \bibinfo{author}{\bibfnamefont{M.}~\bibnamefont{Mark}},
  \bibinfo{author}{\bibfnamefont{A.}~\bibnamefont{Rietzler}},
  \bibinfo{author}{\bibfnamefont{J.}~\bibnamefont{Schindler}},
  \bibinfo{author}{\bibfnamefont{E.}~\bibnamefont{Zupanic}},
  \bibinfo{author}{\bibfnamefont{R.}~\bibnamefont{Grimm}}, \bibnamefont{and}
  \bibinfo{author}{\bibfnamefont{F.}~\bibnamefont{Ferlaino}},
  \bibinfo{journal}{Phys. Rev. A} \textbf{\bibinfo{volume}{85}},
  \bibinfo{pages}{051401} (\bibinfo{year}{2012}).

\bibitem[{\citenamefont{Anderson et~al.}(1995)\citenamefont{Anderson, Ensher,
  Matthews, Wieman, and Cornell}}]{anderson1995observation}
\bibinfo{author}{\bibfnamefont{M.~H.} \bibnamefont{Anderson}},
  \bibinfo{author}{\bibfnamefont{J.~R.} \bibnamefont{Ensher}},
  \bibinfo{author}{\bibfnamefont{M.~R.} \bibnamefont{Matthews}},
  \bibinfo{author}{\bibfnamefont{C.~E.} \bibnamefont{Wieman}},
  \bibnamefont{and} \bibinfo{author}{\bibfnamefont{E.~A.}
  \bibnamefont{Cornell}}, \bibinfo{journal}{Science}
  \textbf{\bibinfo{volume}{269}}, \bibinfo{pages}{198} (\bibinfo{year}{1995}).

\bibitem[{\citenamefont{Davis et~al.}(1995)\citenamefont{Davis, Mewes, Andrews,
  Van~Druten, Durfee, Kurn, and Ketterle}}]{davis1995bose}
\bibinfo{author}{\bibfnamefont{K.}~\bibnamefont{Davis}},
  \bibinfo{author}{\bibfnamefont{M.}~\bibnamefont{Mewes}},
  \bibinfo{author}{\bibfnamefont{M.}~\bibnamefont{Andrews}},
  \bibinfo{author}{\bibfnamefont{N.}~\bibnamefont{Van~Druten}},
  \bibinfo{author}{\bibfnamefont{D.}~\bibnamefont{Durfee}},
  \bibinfo{author}{\bibfnamefont{D.}~\bibnamefont{Kurn}}, \bibnamefont{and}
  \bibinfo{author}{\bibfnamefont{W.}~\bibnamefont{Ketterle}},
  \bibinfo{journal}{Phys. Rev. Lett.} \textbf{\bibinfo{volume}{75}},
  \bibinfo{pages}{3969} (\bibinfo{year}{1995}).

\bibitem[{\citenamefont{Inguscio et~al.}(2008)\citenamefont{Inguscio, Ketterle,
  and Salomon}}]{Inguscio2006ufg}
\bibinfo{editor}{\bibfnamefont{M.}~\bibnamefont{Inguscio}},
  \bibinfo{editor}{\bibfnamefont{W.}~\bibnamefont{Ketterle}}, \bibnamefont{and}
  \bibinfo{editor}{\bibfnamefont{C.}~\bibnamefont{Salomon}}, eds.,
  \emph{\bibinfo{title}{Ultra-cold Fermi Gases}} (\bibinfo{publisher}{IOS
  Press, Amsterdam}, \bibinfo{year}{2008}), \bibinfo{note}{{P}roceedings of the
  International School of Physics ``Enrico Fermi'', Course CLXIV, Varenna,
  20-30 June 2006}.

\bibitem[{\citenamefont{Taie et~al.}(2010)\citenamefont{Taie, Takasu, Sugawa,
  Yamazaki, Tsujimoto, Murakami, and Takahashi}}]{taie2010realization}
\bibinfo{author}{\bibfnamefont{S.}~\bibnamefont{Taie}},
  \bibinfo{author}{\bibfnamefont{Y.}~\bibnamefont{Takasu}},
  \bibinfo{author}{\bibfnamefont{S.}~\bibnamefont{Sugawa}},
  \bibinfo{author}{\bibfnamefont{R.}~\bibnamefont{Yamazaki}},
  \bibinfo{author}{\bibfnamefont{T.}~\bibnamefont{Tsujimoto}},
  \bibinfo{author}{\bibfnamefont{R.}~\bibnamefont{Murakami}}, \bibnamefont{and}
  \bibinfo{author}{\bibfnamefont{Y.}~\bibnamefont{Takahashi}},
  \bibinfo{journal}{Phys. Rev. Lett.} \textbf{\bibinfo{volume}{105}},
  \bibinfo{pages}{190401} (\bibinfo{year}{2010}).

\bibitem[{\citenamefont{Stellmer et~al.}(2013)\citenamefont{Stellmer, Grimm,
  and Schreck}}]{stellmer2013production}
\bibinfo{author}{\bibfnamefont{S.}~\bibnamefont{Stellmer}},
  \bibinfo{author}{\bibfnamefont{R.}~\bibnamefont{Grimm}}, \bibnamefont{and}
  \bibinfo{author}{\bibfnamefont{F.}~\bibnamefont{Schreck}},
  \bibinfo{journal}{Phys. Rev. A} \textbf{\bibinfo{volume}{87}},
  \bibinfo{pages}{013611} (\bibinfo{year}{2013}).

\bibitem[{\citenamefont{Bohn et~al.}(2009)\citenamefont{Bohn, Cavagnero, and
  Ticknor}}]{bohn2009quasi-universal}
\bibinfo{author}{\bibfnamefont{J.~L.} \bibnamefont{Bohn}},
  \bibinfo{author}{\bibfnamefont{M.}~\bibnamefont{Cavagnero}},
  \bibnamefont{and} \bibinfo{author}{\bibfnamefont{C.}~\bibnamefont{Ticknor}},
  \bibinfo{journal}{New J. Phys.} \textbf{\bibinfo{volume}{11}},
  \bibinfo{pages}{055039} (\bibinfo{year}{2009}).

\bibitem[{not({\natexlab{c}})}]{note4}
\bibinfo{note}{For estimating $v$, $T'$ is determined by fitting a Gaussian
  distribution to the observed TOF absorption images over the entire
  temperature range such that $v$ includes the effect of a momentum spread by
  the Fermi energy, as performed in Ref.\,\cite{demarco1999measurement}.}

\bibitem[{\citenamefont{Monroe et~al.}(1993)\citenamefont{Monroe, Cornell,
  Sackett, Myatt, and Wieman}}]{monroe1993measurement}
\bibinfo{author}{\bibfnamefont{C.~R.} \bibnamefont{Monroe}},
  \bibinfo{author}{\bibfnamefont{E.~A.} \bibnamefont{Cornell}},
  \bibinfo{author}{\bibfnamefont{C.~A.} \bibnamefont{Sackett}},
  \bibinfo{author}{\bibfnamefont{C.~J.} \bibnamefont{Myatt}}, \bibnamefont{and}
  \bibinfo{author}{\bibfnamefont{C.~E.} \bibnamefont{Wieman}},
  \bibinfo{journal}{Phys. Rev. Lett.} \textbf{\bibinfo{volume}{70}},
  \bibinfo{pages}{414} (\bibinfo{year}{1993}).

\bibitem[{\citenamefont{DeMarco et~al.}(1999)\citenamefont{DeMarco, Bohn,
  Burke, Holland, and Jin}}]{demarco1999measurement}
\bibinfo{author}{\bibfnamefont{B.}~\bibnamefont{DeMarco}},
  \bibinfo{author}{\bibfnamefont{J.~L.} \bibnamefont{Bohn}},
  \bibinfo{author}{\bibfnamefont{J.~P.} \bibnamefont{Burke}},
  \bibinfo{author}{\bibfnamefont{M.}~\bibnamefont{Holland}}, \bibnamefont{and}
  \bibinfo{author}{\bibfnamefont{D.~S.} \bibnamefont{Jin}},
  \bibinfo{journal}{Phys. Rev. Lett.} \textbf{\bibinfo{volume}{82}},
  \bibinfo{pages}{4208} (\bibinfo{year}{1999}).

\bibitem[{\citenamefont{Riedl et~al.}(2008)\citenamefont{Riedl, {S\'{a}nchez
  Guajardo}, Kohstall, Altmeyer, Wright, {Hecker Denschlag}, Grimm, Bruun, and
  Smith}}]{Riedl2008coo}
\bibinfo{author}{\bibfnamefont{S.}~\bibnamefont{Riedl}},
  \bibinfo{author}{\bibfnamefont{E.~R.} \bibnamefont{{S\'{a}nchez Guajardo}}},
  \bibinfo{author}{\bibfnamefont{C.}~\bibnamefont{Kohstall}},
  \bibinfo{author}{\bibfnamefont{A.}~\bibnamefont{Altmeyer}},
  \bibinfo{author}{\bibfnamefont{M.~J.} \bibnamefont{Wright}},
  \bibinfo{author}{\bibfnamefont{J.}~\bibnamefont{{Hecker Denschlag}}},
  \bibinfo{author}{\bibfnamefont{R.}~\bibnamefont{Grimm}},
  \bibinfo{author}{\bibfnamefont{G.~M.} \bibnamefont{Bruun}}, \bibnamefont{and}
  \bibinfo{author}{\bibfnamefont{H.}~\bibnamefont{Smith}},
  \bibinfo{journal}{Phys. Rev. A} \textbf{\bibinfo{volume}{78}},
  \bibinfo{eid}{053609} (\bibinfo{year}{2008}).

\bibitem[{\citenamefont{Massignan et~al.}(2005)\citenamefont{Massignan, Bruun,
  and Smith}}]{Massignan2005vra}
\bibinfo{author}{\bibfnamefont{P.}~\bibnamefont{Massignan}},
  \bibinfo{author}{\bibfnamefont{G.~M.} \bibnamefont{Bruun}}, \bibnamefont{and}
  \bibinfo{author}{\bibfnamefont{H.}~\bibnamefont{Smith}},
  \bibinfo{journal}{Phys. Rev. A} \textbf{\bibinfo{volume}{71}},
  \bibinfo{eid}{033607} (\bibinfo{year}{2005}).

\bibitem[{\citenamefont{Bruun and Smith}(2005)}]{Bruun2005vat}
\bibinfo{author}{\bibfnamefont{G.~M.} \bibnamefont{Bruun}} \bibnamefont{and}
  \bibinfo{author}{\bibfnamefont{H.}~\bibnamefont{Smith}},
  \bibinfo{journal}{Phys. Rev. A} \textbf{\bibinfo{volume}{72}},
  \bibinfo{eid}{043605} (\bibinfo{year}{2005}).

\bibitem[{\citenamefont{Pethick and Smith}(2002)}]{Pethick2002book}
\bibinfo{author}{\bibfnamefont{C.~J.} \bibnamefont{Pethick}} \bibnamefont{and}
  \bibinfo{author}{\bibfnamefont{H.}~\bibnamefont{Smith}},
  \emph{\bibinfo{title}{Bose-Einstein condensation in dilute gases}}
  (\bibinfo{publisher}{Cambridge University Press}, \bibinfo{year}{2002}).

\end{thebibliography}

\end{document}